# Tracing Optimization for Performance Modeling and Regression Detection

KAVEH SHAHEDI, Polytechnique Montréal, Canada
HENG LI, Polytechnique Montréal, Canada
MAXIME LAMOTHE, Polytechnique Montréal, Canada
FOUTSE KHOMH, Polytechnique Montréal, Canada

Software performance modeling plays a crucial role in developing and maintaining software systems. A performance model analytically describes the relationship between the performance of a system and its runtime activities. This process typically examines various aspects of a system's runtime behavior, such as the execution frequency of functions or methods, to forecast performance metrics like program execution time. By using performance models, developers can predict expected performance and thereby effectively identify and address unexpected performance regressions when actual performance deviates from the model's predictions. One common and precise method for capturing performance behavior is software tracing, which involves instrumenting the execution of a program, either at the kernel level (e.g., system calls) or application level (e.g., function calls). However, due to the nature of tracing, it can be highly resource-intensive, making it impractical for production environments where resources are limited. In this work, we propose statistical approaches to reduce tracing overhead by identifying and excluding performance-insensitive code regions, particularly application-level functions, from tracing while still building accurate performance models that can capture performance degradations. By selecting an optimal set of functions to be traced, we can construct optimized performance models that achieve an $R^2$ score of up to 99% and, in some cases, outperform full tracing models (i.e., models using non-optimized tracing data), while significantly reducing the tracing overhead by more than 80% in most cases. Our optimized performance models can also capture performance regressions in our studied programs effectively, demonstrating their usefulness in real-world scenarios. Finally, our approach is fully automated, making it ready to be used in production environments with minimal human effort.

## 1 INTRODUCTION

Software performance (e.g., execution time or resource usage) plays a critical role in the overall quality of a software system [12, 28, 54, 55]. Sub-optimal performance can lead to significant consequences for the system operators and users, such as unnecessary operation overhead or jeopardized user experience [17, 61, 64]. Performance modeling has been widely used for various performance assurance activities, such as performance regression detection [8, 9, 43, 51, 60], performance bug localization [16, 51], performance prediction [34, 58, 63], scalability analysis [19, 37], capacity planning [6, 14], and resource allocation [10, 31]. Essentially, a performance model describes the relationship between a

Authors' addresses: Kaveh Shahedi, kaveh.shahedi@polymtl.ca, Polytechnique Montréal, Montréal, QC, Canada; Heng Li, heng.li@polymtl.ca, Polytechnique Montréal, Montréal, QC, Canada; Maxime Lamothe, maxime.lamothe@polymtl.ca, Polytechnique Montréal, Montréal, QC, Canada; Foutse Khomh, foutse.khomh@polymtl.ca, Polytechnique Montréal, Montréal, QC, Canada.

performance metric $P$ (e.g., execution time) and the runtime activities ($X$) of a system: $P = f(X)$ [12, 55]. Performance metrics can be measured efficiently during system execution (e.g., by using system monitoring tools such as *pidstat*).

To this end, tracing is a common way of capturing the detailed runtime activities of a software system. Basically, it is an advanced logging technique where a tracing tool establishes a framework for efficient and customizable logging, allowing developers to monitor the progression of operations within a program by observing events at both the application level (e.g., function calls within the program) and at the kernel level (e.g., operating system calls) [1, 2, 21, 26, 30, 39]. Debugging, performance analysis, and system optimization are a few of the multiple possible usages of software tracing [6, 24, 33]. Prior work has used tracing to build performance models [6, 13, 35, 41, 59]. However, the cost of capturing and storing detailed information about a system's execution can lead to significant overhead, increasing the application's computational costs and resource usage [30]. However, considering tracing-based performance modeling, not all code regions (i.e., application-level functions) in a program significantly contribute to the system's overall performance [6, 33, 59], and applying tracing on those regions may not enhance the precision of the performance model.

In this work, we propose statistical approaches that considerably reduce the overhead of application-level tracing while maintaining sufficient trace information to build accurate performance models. Our intuition is that not all functions in a program significantly affect the overall performance [6, 33, 59], and tracing these functions may not contribute to the precision of the resulting performance model. Thus, we **hypothesize** that we can remove these performance-insensitive functions from tracing, thereby decreasing the overhead of tracing considerably. We propose 9 dynamic approaches in addition to one static approach to identify performance-insensitive functions within a program. In the dynamics, a function is deemed **performance-insensitive** and thus eligible for **removal from tracing** if: (1) it consistently demonstrates stable performance across various inputs, as evidenced by either low *Shannon's Entropy* or *Coefficient of Variation (CoV)* in its performance distribution; or (2) it does not provide significant information to the performance model, as indicated by a high correlation with another function's performance or low contribution impact on the performance model (i.e., *p-value* $< 0.05$). In the static approach, we evaluate the static code features of each function (e.g., lines of code, number of loops) and assign a score to determine whether it falls into the category of performance-sensitive or performance-insensitive functions.

Using the optimized trace data (i.e., tracing only performance-sensitive functions), we build **performance models** to describe the relationship between the overall performance of the system (i.e., execution time) and the detailed runtime activities recorded in the trace (i.e., call frequencies of the functions). We use black-box machine learning performance models to enable us to predict a performance metric of the system based on the observed runtime activities of the system (e.g., function calls recorded in the trace data) without requiring domain knowledge [22, 29, 43]. By building various optimized performance models, each leveraging the performance-sensitive functions of a pruning approach (see 3.2.2), we assess their prediction accuracy to determine the impact of tracing only performance-sensitive functions. For each pruning approach (i.e., 9 dynamic and one static), we build multiple performance models, each based on a specific machine learning model, to assess which pruning approach yields the best set of performance-sensitive functions. We also show that the choice of machine learning model is crucial for each program and each pruning approach.

In addition to evaluating the efficacy of our tracing optimization approaches and the efficiency of the optimized performance models, we also evaluate the effectiveness of the resulting performance models in a more practical setting: model-based **performance regression detection**. In particular, regression detection is the process of identifying and diagnosing undesirable performance drops in comparison to the performance of a previous version of the software [27, 47]. Regression detection techniques can identify specific areas where performance has decreased and provide insights into the root causes, facilitating quick correction and ensuring software stability and reliability. Multiple investigations have focused on detecting performance regressions within systems using various performance models [8, 9, 43, 51, 60]. As performance models capture the relationship between the performance of a software system and its runtime activities with the goal of predicting specific performance metrics (e.g., execution time, resource utilization, etc.), unanticipated performance fluctuations (e.g., a longer execution time for the same amount of runtime activities) can result in errors in the model's predictions, thereby suggesting performance regressions. Employing performance models can assist developers in identifying and pinpointing problematic performance-related segments of a system (e.g., code regions), enabling early-stage diagnosis and resolution. In order to assess the capability of our optimized performance models in detecting performance regressions, we manually inject performance regressions into the studied programs and test the effectiveness of the models in identifying them [4, 43, 51].

We assess our work by answering the three research questions:

- *RQ1: How does tracing only performance-sensitive functions impact the tracing overhead?* Not all functions within a program significantly influence the overall performance and, accordingly, do not provide valuable insights to be used in the performance models. We consider a comprehensive set of trace pruning approaches that are aimed at pinpointing the performance-insensitive functions. Then, we evaluate the impact of each pruning approach on mitigating the system's tracing overhead (e.g., execution time or storage usage). We show that only a limited number of functions (i.e., 5.64% to 21.93%) are performance-sensitive, and removing the rest from tracing can reduce the overhead of tracing significantly.
- *RQ2: How well can we build performance models with only performance-sensitive functions?* There is a possibility of accuracy loss when building an optimized performance model by taking only the performance-sensitive functions into account (i.e., due to loss of runtime activity information). Therefore, we evaluate the accuracy of the performance models that are built using the optimized tracing information (i.e., only considering the performance-sensitive function). Also, we indicate which pruning approach leads to a more precise performance model for each studied program. The results indicate that our optimized performance models can achieve an $R^2$ score of up to 99%, and in some cases, they can outperform the non-optimized performance models.
- *RQ3: How effective are the optimized performance models in detecting performance regressions?* Performance regression detection is one of the main practical usages of performance models. To evaluate the effectiveness of our optimized models in this regard, we manually inject performance regressions into the programs and assess how well the optimized performance models can detect them. We demonstrate that our optimized performance models can effectively distinguish between program versions with and without performance regression.

Our work makes several main contributions:

(1) *We propose multiple tracing optimization approaches (i.e., both dynamic and static) that are capable of substantially reducing tracing overhead.*
(2) *We present an efficient framework for tracing-based performance modeling and regression detection based on our proposed tracing optimization approaches.*
(3) *We conduct a comprehensive empirical evaluation of our tracing optimizations and performance modelings to assess the effectiveness of each approach.*

The paper is structured as follows: Section 3 provides detailed information about our methodology, including tracing optimization methodologies, performance modeling procedures, and regression detection approaches. In Section 4, we assess the effectiveness of our implemented approach. Section 2 discusses related works, while Section 5 examines potential threats to the validity of our work. Section 6 concludes our work.

## 2 RELATED WORK

### 2.1 Tracing Overhead Reduction

Mitigating the overhead of tracing is an ongoing challenge, as it requires finding a balance between the information gained and the additional work the tracing imposes on the system. Previous research has explored different methods to reduce this overhead, using both static and dynamic analyses. For instance, in a study by J. Mußler et al. [45], they took a static approach to identify the most valuable functions in a system based on metrics like lines of code, complexity, and loops. They then excluded the less important functions from direct instrumentation, which led to a significant reduction in overhead as a large portion of functions were removed from the studied programs. In another research effort by Lehr et al. [42], a combination of static and dynamic analyses was used to pinpoint the most critical code sections. They employed adjustable filters for instrumentation, allowing them to control overhead while maximizing the information obtained. Building upon this work [41], they expanded their program analyses and used dynamic results from the applications' performance metrics to create empirical performance models. Similarly, N. Kumar et al. [40] proposed an approach to lower instrumentation overhead. They achieved this by 1) reducing the number of tracepoints in the program, 2) minimizing the cost of these tracepoints, and 3) optimizing the cost of the tracepoints' data. Their approach aimed to apply these changes to the program while preserving its original tracing semantics.

Following the prior works, we aimed to refine the number of functions to be traced by observing the performance behaviors of the application-level function to keep the most performance-sensitive functions in the tracing procedure.

### 2.2 Performance Modeling and Regression Detection.

Performance modeling is a widely investigated topic that generally aims to produce consistent and accurate performance models in the production stage of applications. In a study by L. Liao et al. [44], black-box models were employed, which leveraged system performance metrics, such as CPU utilization, to estimate how programs behaved in terms of performance. When these models detected a performance degradation in the system, the approach traced back to historical code changes in the program's development process to identify the root cause of the regression. In a related study [43], the authors used machine learning techniques to determine the connection between problematic requests that led to performance regressions in the system. They constructed black-box performance models to spot variations in system performance and, subsequently, used machine learning, specifically linear regression models, to link these regressions to the corresponding requests, thus identifying the root causes. Another approach, as seen in the work by KC. Foo et al. [27], aimed to build ensemble performance models. These models collected data from various system sections and relied on historical data to detect performance regressions. This approach was effective since it does not depend on a specific configuration, making it applicable to a wide range of heterogeneous systems. In [6], the authors employed system tracing techniques to assess the performance of large-scale cloud applications that experience significant variations in system load in production. They introduce a coordinated burst tracing method designed to comprehensively capture activities across all system layers. This data is then used to analyze the system's performance.

As a combination of the prior works regarding performance modeling and regression detection, in this study, we leverage program's trace data in order to build accurate performance models that are capable of detecting performance regressions in the system. Also, mitigating the overhead of tracing was the primary objective, making the final optimized model more useful in production due to less overhead on the system.

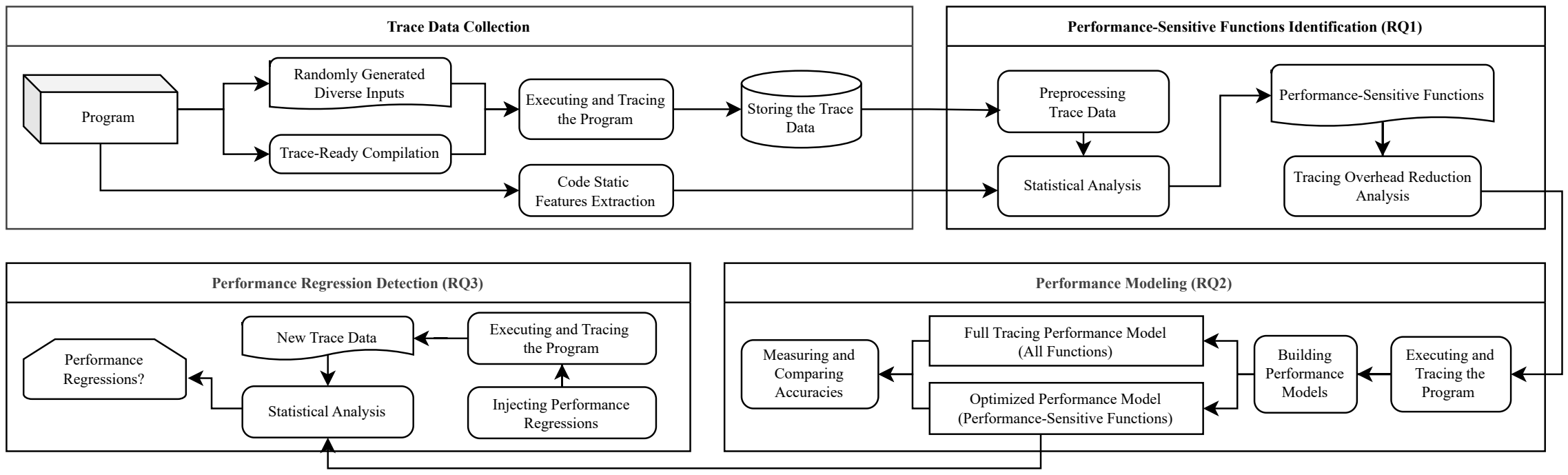

Fig. 1. The overview of determining performance-sensitive functions, optimized performance modeling, and regression detection.

## 3 METHODOLOGY

In this section, we present the methodology of our framework for trace optimization, as well as performance modeling and regression detection that are built upon the optimized tracing. Figure 1 shows an overview of our methodology. Considering a target program (i.e., a program that we want to reduce its tracing overhead and build performance models for), we first leverage *uftrace* [39] to trace all functions within the program by executing it with a set of different inputs and collect the trace data along with performance metrics (i.e., the *trace data collection* step). Then, we perform statistical analysis on the trace data to identify the functions that have the most significant impact on the system's performance (i.e., the *performance-sensitive functions identification* step). Then, we refine the tracing to focus on the performance-sensitive functions (i.e., to reduce performance and resources overhead) and leverage the reduced set of tracing to build performance models (i.e., the *performance modeling* step). Finally, we leverage the performance models to detect injected performance regressions (i.e., the *performance regression detection* step), which demonstrates a practical application scenario of the performance models.

### 3.1 Trace Data Collection

We describe our subject program selection, their input generation, our tracing method, and trace data collection below.

*3.1.1* ***Subject Programs***. The primary methodology in this research involves leveraging software traces to observe the behavior of individual functions within a program and assess the value of the information they provide when building a performance model. To achieve this, we focus on three widely recognized benchmark suites: *SPEC CPU 2017*[15], *SU2*[23], and *PARSEC 3.0* [11].

- *SPEC CPU 2017* is a diverse collection of benchmark programs designed to evaluate CPU performance across various system configurations.
- *SU2* comprises open-source C++ applications for solving partial differential equations and performing optimization constrained by these equations.
- *PARSEC 3.0* is a suite of programs aimed at measuring the performance of multiprocessor systems.

These benchmarks are frequently used in performance studies [41, 42, 45, 48] and provide a broad spectrum of scenarios to evaluate our approach. Our study focuses exclusively on programs developed in C or C++, as these languages are commonly used for performance-critical applications [15, 23, 32]. Additionally, *uftrace* [39], the tracing tool used in this research, primarily supports these languages and their associated compilers. To generate new randomized inputs for

Table 1. Specifications of the studied programs

| | Program | | | | |
|---|---|---|---|---|---|
| **Metric** | **SU2** | **638.imagick** | **657.xz** | **631.deepsjeng** | **freqmine** |
| Programming Language | C/C++ | C | C | C++ | C++ |
| KLOC | 900 | 259 | 33 | 10 | 4 |
| Number of Functions | >15k | 2,792 | 419 | 123 | 56 |

*Note:* KLOC refers to the Lines of Code/1000.

program execution (see Section 3.1.2), we selected programs with accessible sample inputs. This criterion was essential to ensure the generation of diverse and effective test cases. Furthermore, we considered static code features such as the number of functions, lines of code, and input size to ensure a diverse range of complexity and program size in our study. This diversity is critical to demonstrating that our methodology is applicable across various types of programs, not just those of a specific size or complexity.

Consequently, we selected five benchmark programs representing a spectrum of code complexity and size: *631.deepsjeng*, *638.imagick*, and *657.xz* from *SPEC CPU 2017*, *SU2*, and *freqmine* from *PARSEC 3.0*. The specifications of the aforementioned programs are shown in Table 1.

*3.1.2* ***Input Generation***. Each of our selected benchmark programs includes sample inputs, but these are generally insufficient to cover most of the program's call paths. Since our goal is to observe fluctuations in the execution of program functions, we need inputs that maximize code coverage by invoking these functions in as many scenarios as possible. To achieve this, we conduct a manual analysis of the input structures, focusing on parameters that can be mathematically randomized, such as numeric or enumerated inputs. Parameters requiring specialized domain knowledge, such as binary inputs, are excluded from this process.

We then automate the generation of random values for the selected parameters by sampling from a wide range of original values typically found in the program's sample inputs. This range is extended to include values from both lower and significantly higher magnitudes. For example, in the case of *631.deepsjeng*, two key parameters are considered: the positions of chess pieces in FEN format and the depth of analysis the program should perform on these positions. To generate new random inputs, we create combinations of chess positions that vary from simple, incomplete scenarios to complex, advanced configurations. Similarly, for the depth parameter, we generate values ranging from low to very high, thereby imposing different levels of computational demand on the program. For other programs, we identify and randomize parameters that fit this approach.

Once we generate a broad set of inputs, we randomly sample a subset (i.e., 2,500 samples) for use in further steps. This sampling strategy ensures a diverse set of inputs, increasing the likelihood of covering different call paths within the programs. The scripts used for input generation are available in our replication package (see Section 7).

*3.1.3* ***Tracing and Compilation***. All the programs used in this study are implemented in C and C++. Therefore, to enable user-space level tracing (i.e., program's functions), we employ *uftrace* [39]. We chose *uftrace* due to its low overhead and efficiency for tracing C and C++ programs, as well as its straightforward usage regarding collecting performance trace data of the programs' functions. To prepare the programs for tracing, they must be compiled with specific profiling flags(i.e., *-pg* or *-finstrument-functions*), which allows us to trace the programs during their execution. Our tracing process solely concentrates on the user-space level, specifically tracing the functions within the programs. Consequently, kernel-level tracing and library calls are omitted from the tracing procedure since we only need information from the functions of the programs. Once the programs are trace-ready, we must then run the programs to obtain traces. We use the following command structure to run the *uftrace* for tracing the studied programs:

```
uftrace record --time --no-libcall <PROGRAM-EXECUTION-COMMAND>
```

*3.1.4* ***Executing the Trace-Ready Program.*** Once the trace-ready programs are prepared with their respective inputs, we proceed to trace their individual executions and store the trace data for subsequent analysis. The following is a list of information that we collect from each execution:

- Program Execution Time: The performance model is aimed to predict specific response variables based on the observed trace data. In this work, we consider the program execution time to be the variable responsible. Program execution time shows the direct interaction of the program with the system, and any unexpected variation in it will impact the system's flows and processes.
- Function Performance Information: Using *uftrace*, we capture two key metrics for each function: *total self-execution time*, which helps identify performance-sensitive functions, and *number of calls*, which serves as an explanatory variable in our performance models.
  - We use self-execution times in our statistical analyses to observe fluctuations in function behavior, particularly measuring how much these times vary across multiple executions. This variation helps us identify performance-sensitive functions. If a function's fluctuation approaches zero after many executions, it suggests that the function is unaffected by the program's inputs and consistently incurs the same performance cost. Such functions, providing little new information to a performance model, can often be modeled using constant values and may be pruned from the tracing process. In contrast, if a function executes differently based on each input, its fluctuation in behavior will be higher. Functions with high fluctuations in their behavior are thus particularly interesting for performance models.
  - For building performance models, we use the call frequencies of functions as explanatory features to predict the program's total execution time. The choice of call frequencies over self-execution times aims to avoid potential bias in the models. If regressions occur, using execution times as variables could skew the model's predictions, as the explanatory variables might be directly influenced by the regression. In contrast, call frequencies remain unaffected by regressions, offering a more stable and reliable basis for accurate prediction.
- Trace Data Characteristics: In addition, we gather other details, including trace overheads in terms of storage usage, which are used for overhead investigations and comparisons.

### 3.2 Performance-Sensitive Functions Identification

Once a program's trace data is stored, we analyze the performance distribution of each function. This procedure aims to evaluate the information that each function produces during different program runs. In order to accomplish this, we compute several statistical measures (see 3.2.2) for each function and divide the functions into two categories: performance-sensitive and performance-insensitive. By categorizing the functions according to their sensitivity to performance, we collect necessary information for subsequent performance model optimizations and adjustments.

*3.2.1* ***Preprocessing Trace Data.*** To analyze the behavior of functions, we focus on two key metrics: *total self-execution time* and *number of calls*. These metrics reflect a function's behavior based on the given inputs, independent of the time taken by its child functions to complete. However, during program execution, unexpected system interrupts may occur, leading to variations in function behavior. To address this, we sort the data for each function and exclude outliers based on the criterion *mean ± 3∗std*. This approach effectively filters out biased data while maintaining high accuracy in our calculations. By applying this outlier removal technique, we obtain reliable insights into function behavior, which aids in identifying performance-sensitive functions. Additionally, we filter out functions that do not provide significant information for our analysis by excluding those with nearly constant call frequencies. Since our performance models use the number of function calls as explanatory features, functions with minimal variation in their call frequencies contribute little to model accuracy. Specifically, we exclude functions whose number of unique call frequencies falls below the top $99^{th}$ percentile of program executions. For example, if a program is executed 1,000 times and a function has fewer than 10 unique call frequencies (e.g., it was executed 2 times in 975 runs, 5 times in 15 runs, and 4 times in 10 runs), we exclude it from our analysis. Including such functions would result in many identical data points, which would negatively impact the accuracy of the performance models.

#### 3.2.2 ***Criteria for Identifying Performance-Sensitive Functions.*** We consider both dynamic (based on runtime information) and static (based solely on static code) approaches to identify performance-sensitive functions.

- Dynamic Approaches: A function is deemed *performance-insensitive* and eligible for removal from tracing if:
  - Consistent Performance: The function consistently performs similarly across different inputs. We assess this using *Shannon's Entropy* and *Coefficient of Variation (CoV)*.
  - Insignificant Contribution to Performance Models: The function contributes insignificantly to the performance model. We evaluate this by analyzing the *correlation* between function performance metrics, and also, by assessing each function's *Feature Significance* within a linear regression-based model.

  Additionally, we consider combinations of these metrics to ensure a more robust analysis.
- Static Approach: For each function, we compute multiple performance-related static code features and assign a static performance sensitivity score (*StaPerfSens*). Functions with a *StaPerfSens* score below a certain threshold are considered performance-insensitive and are excluded from tracing.

Below is a detailed description for each pruning methodology. We indicate the dynamic approaches with *(D)*, and the static approach with *(S)*.

Shannon's Entropy (D): This metric expresses the degree of uncertainty or information included in a dataset; a larger value denotes more randomness, while a lower value indicates more predictability [52]. By calculating this metric for each function, we can evaluate the extent of execution uncertainty and identify functions with higher uncertainty regarding execution time. Particularly, as a consequence of their larger fluctuations, functions with higher entropy values are more likely to impact the system's performance significantly. On the other hand, functions with lower values of entropy show more stable and predictable behavior. In the collected information, as the execution times are continuous while *Shannon's Entropy* is designed for discrete problems, we normalize all the values and group them into fixed-size bins. To determine the appropriate fixed-size bin, we calculate individual bin sizes using *Scott's Rule* [50] for each function and then compute the average bin size, which serves as the final bin size used for the analysis. By binning the entire dataset, we calculate the *Shannon's Entropy* of the distribution of each function's execution time values using the following formula:

$$H(X) = -\sum_{i=1}^{n} p(x_i) \log_b(p(x_i))$$

where $x_i$ corresponds to each unique value in the distribution of the functions' execution time, and $p(x_i)$ represents the probability of observing each unique value $x_i$ in the distribution.

Coefficient of Variation (CoV) (D): It indicates the relative variability of a dataset by comparing the standard deviation to the mean, allowing for the comparison of datasets with different scales. This metric has been previously used in studies to capture the fluctuation of execution times [7, 38].

$$CV = \left( \frac{\sqrt{\frac{1}{N}\sum_{i=1}^{N}(x_i - \mu)^2}}{\frac{1}{N}\sum_{i=1}^{N} x_i} \right)$$

In this scenario, $x_i$ corresponds to each value in the distribution of the functions' execution time, $N$ represents the number of the values in the distribution, and $\mu$ is the average of the data points in the distribution. If we simplify the formula, the numerator represents the *Standard Deviation* of the distribution, and the denominator shows the mean (average) of the values in the distribution. Similar to *Shannon's Entropy*, functions with higher *CoV* are more likely to impact performance due to their greater execution time variations. Our investigations revealed that the *Entropy* of functions' executions might differ considerably from their *CoV*.

Union of Shannon's Entropy and CoV (D): After preliminary investigations, we observe variations between the candidate performance-sensitive functions obtained from the *Entropy* and *CoV* criteria since fluctuation and uncertainty in a function's behavior do not necessarily align. Consider an example where two functions are each executed 100 times during a single program run. Both functions have self-execution times that fluctuate around a specific point, say $100ns$. However, the nature of these fluctuations differs between the two functions. For the first function, the amount of fluctuations is high but their magnitude is relatively small, around $0.5ns$. Despite this, the function exhibits high *Entropy* because its behavior is unpredictable (i.e., executes differently each time), even though the fluctuations are minor. The *CoV*, however, is low due to the stability of these minor fluctuations. In contrast, the second function has a more consistent behavior in terms of the number of distinct execution times—say it exhibits ten distinct self-execution times. However, the range of these times is much broader, for instance, around $50ns$. This results in low *Entropy* (due to fewer distinct states) but a high *CoV*, reflecting the larger spread in its execution times. This example illustrates that high *Entropy* does not necessarily correspond to a high *CoV* and vice versa. Therefore, we create a union set of the candidate functions from both criteria for further analysis, aiming to assess if combining the characteristics of these two metrics can actually result in a more accurate performance model, since they provide both fluctuations and uncertainty of the functions' executions.

Performance Correlations (D): In a program, the performance behavior of some functions may be highly correlated, causing redundancy for the potential performance models and, accordingly, negatively impacting their accuracy. Thus, we exclude highly correlated functions and retain a subset for tracing and model building. We employ the *Spearman Correlation* [56] to compute the correlations between the performance distributions of each pair of the functions. This choice is made because the performance data does not follow a normal distribution (we performed the *Shapiro-Wilk Test* [53] to determine the normality of our data). We use this approach independently (i.e., having a set of performance-sensitive functions retained by only applying correlation removal) and its combination with other metrics, including *Shannon's Entropy* and *CoV* (i.e., removing the highly correlated functions in the set of performance-sensitive functions resulting from *Shannon's Entropy* and *CoV*).

Feature Significance (D): In this criterion, we aim to identify functions that significantly influence the performance model. Unlike our prior pruning methodologies, such as *Shannon's Entropy*, *CoV*, and *Performance Correlations*, we try to identify performance-sensitive functions without employing specific statistical techniques. Indeed, we train a straightforward linear regression model using all the functions' call frequencies as the model's explanatory features and the program's total execution time as the prediction target. Then, we assess the coefficients of the features within the model and select those that are statistically significant, indicated by a *p-value* (the result of a *F-Test*) below 0.05. These corresponding functions are considered performance-sensitive functions since they exert a noteworthy impact on the performance model.

Union of Dynamic Pruning Methodologies (D)[1]: *Shannon's Entropy* and *CoV* pertain to sensitivity analysis of the functions' performance behaviors (i.e., execution time), *Performance Correlations* endeavor to eliminate highly correlated features regardless of influence on the performance models, and *Feature Significance* is a supervised approach that identifies which features contribute the most to the performance models. The mentioned dynamic approaches consider performance-sensitive functions from two mutually complementary perspectives: performance stability of individual functions (*Shannon's Entropy* and *CoV*) and potential contribution to a performance model (*Performance Correlations* and *Feature Significance*). Thus, we also consider all these criteria collectively by making a union set of the candidate performance-sensitive functions from these individual criteria as a distinct pruning criterion to evaluate the efficacy of their combined utilization.

Static Performance Sensitivity (StaPerfSens) Score (S): In addition to dynamic approaches, we also propose a static approach inspired by prior research on tracing overhead refinement [45], which investigates the use of individual code features (e.g., lines of code) for tracing overhead reduction (e.g., by only tracing long functions). Here, we propose a static performance sensitivity (*StaPerfSens*) score that aggregates these individual code features into a single score. For each program, we compute various static code features, including *lines of code*, *number of loops and nested loops*, *function calls*, *function recursion status*, *number of branches*, and *number of function parameters*. We then rank the functions by their *StaPerfSens* score. Subsequently, employing a one-dimensional clustering technique (see 3.2.3), we identify the performance-sensitive functions for tracing.

*3.2.3* ***Automatically Finding the Threshold for the Rank-Based Metrics***. *Shannon's Entropy*, *CoV*, and *StaPerfSens* pruning methods tend to rank the functions, and accordingly, after measuring them, we have three sets of functions that are sorted based on their *Entropy*, *Entropy*, or *StaPerfSens*. In order to cluster them into two groups (i.e., performance-sensitive and performance-insensitive), we use an automated approach to find a statistical threshold to separate them (i.e., using *Ckmeans.1d.dp algorithm* [57], a one-dimensional clustering method). Specifically, we employ a two-dimensional representation to depict the sorted functions, where the Y values represent the function's metrics (e.g., *Entropy*, *CoV*, or *StaPerfSens*), and the X values correspond to the function's indexes in the sorted list (a numeric array). Next, the Y values are smoothed using the *loess* [3] function, and the first derivative of the Y values with respect to the X values is calculated. We use the first derivative to capture the trend of the Y values: a threshold can be made at the point where the decrease of the Y values starts to slow down significantly (i.e., where the absolute value of the first derivative becomes significantly smaller). Smoothing ensures the stability of the derivative values. Prior work uses a similar approach to automatically determine a threshold [20]. Subsequently, we use the *Ckmeans.1d.dp algorithm* [57], a one-dimensional clustering method, to identify a breakpoint separating the derivatives into two clusters. These groups categorize the functions as either performance-sensitive or performance-insensitive. Figure 2 demonstrates the overview of automatically determining the threshold to split the functions into two sets. In the right plot, the right-hand side functions are considered performance-sensitive, whereas the left-hand ones are performance-insensitive and are eligible for pruning from tracing.

[1]We don't make a union of the dynamic approaches and the static approach, as the dynamic approaches require to execute the programs with all the functions traced to collect their performance information for identify the performance-sensitive ones, whereas the static approach does not.

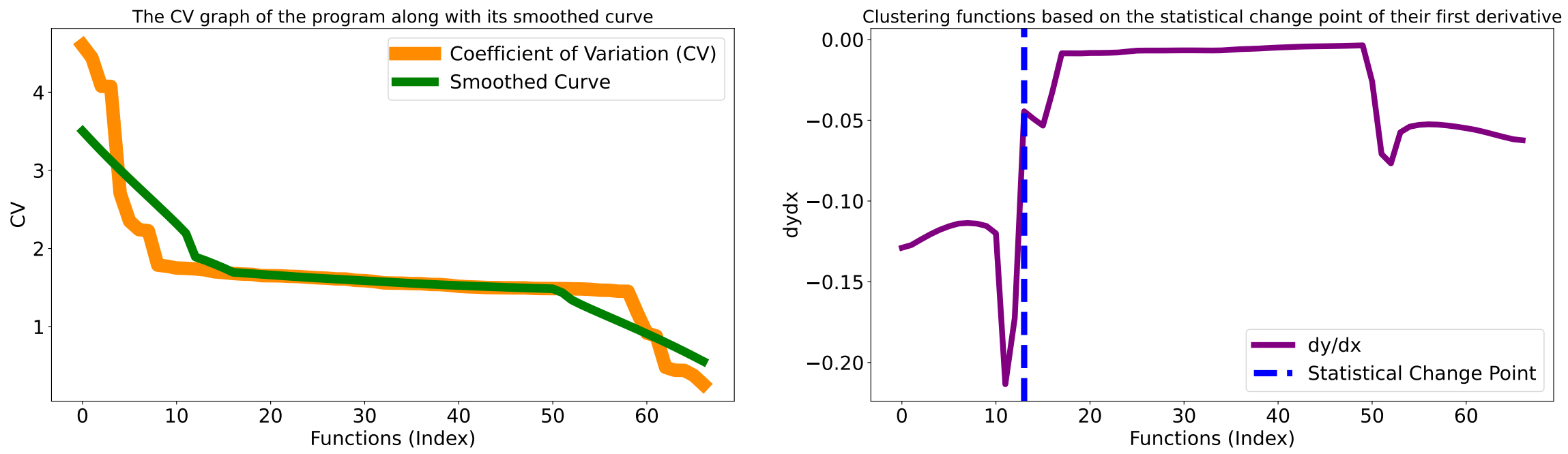

Fig. 2. The *CoV* graph of the *631.deepsjeng* program along with its smoothed curve using *loess*.

### 3.3 Performance Modeling

Black-box machine learning performance models are used to predict a system's performance metric using the system's behavior without having full knowledge of it [22, 29, 43]. Our performance models aim to predict system performance (i.e., execution time) by leveraging trace data (i.e., call frequencies) of the program's functions as explanatory variables.

*3.3.1 Model Training and Evaluation.* We utilized *PyCaret* [5], an open-source Python AutoML library, to automate and streamline the processes of training, testing, and deploying machine learning models. While we initially experimented with manually selecting and analyzing various models, such as Linear Regression and Random Forest, the results obtained through *PyCaret* proved to be more robust and reliable. Specifically, we used *PyCaret's* regression module to build our performance models. Also, since the explanatory features (i.e., functions' call frequencies) were integer values, no additional preprocessing was required. *PyCaret's* built-in comparison feature enabled us to evaluate the effectiveness of different machine learning models for each program and pruning method (e.g., *Entropy*, *CoV*). For model evaluation, we employed 10-fold cross-validation on 80% of the dataset, ensuring a comprehensive assessment of each model's performance. From these results, we selected the top five models which demonstrated the highest accuracy. We then refined these models using *PyCaret's* hyperparameter tuning module. To finalize the evaluation, we tested the tuned models on the remaining 20% of the dataset. The primary evaluation metrics for the performance models were the $R^2$ score, Mean Absolute Error (MAE), and Root Mean Squared Error (RMSE). The $R^2$ score was used to assess how well the model explains the variance in execution time, while MAE measured the average magnitude of errors, and RMSE accounted for larger errors by penalizing them more heavily. Among these, $R^2$ was prioritized as the primary metric in our work for comparing the overall accuracy of the different performance models.

*3.3.2 Baseline: Vanilla (Full-Tracing) Performance Model.* We implement a tracing scenario in which all function events within a program are traced to build a performance model using the complete set of functions. This model, referred to as the "vanilla" or full-tracing performance model, serves as a comprehensive representation of the program's execution. However, full tracing introduces significant system overhead, as it is known to be highly resource-intensive [30, 40, 42, 45]. We establish the vanilla performance model as a baseline to evaluate both the accuracy and the overhead—specifically, the impact on execution time and storage usage—when compared to the optimized performance models.

*3.3.3 Optimized Performance Model.* In contrast to the full-tracing model, the optimized performance model uses the call frequencies of performance-sensitive functions collected from optimized trace data (i.e., tracing only the performance-critical functions) as explanatory variables. Once the performance-sensitive functions are identified for each program, we collect new trace data limited to these functions to assess the accuracy of the optimized model. Because this optimized tracing imposes far less overhead on the system, it can be used in production environments at the commit level to build performance models without significant impact. To create a reliable dataset for training and testing these optimized models, we run the programs with previously unseen inputs. To ensure statistical significance, we calculate a sample size for the new inputs based on a 95% confidence level and a 5% margin of error of the dataset we used to identify performance-sensitive functions (i.e., analysis dataset). Given that this contains 2,500 data points, the required sample size is 333. We then execute the programs with 333 new inputs, collecting 333 corresponding data points for training and testing the performance models.

### 3.4 Performance Regressions Detection

One of the key objectives of performance models is to identify performance regressions occurring in the system [25, 43, 51, 62]. It's essential to recognize that the workload during software system execution is not constant over time. Black-box performance models encapsulate program performance patterns even when the workload changes; when a performance regression occurs, it induces changes in the program's performance behavior. Discrepancies between the performance model's predictions and new observations are indicators for performance regressions.

We aim to determine how using only performance-sensitive functions can affect the ability of the optimized performance model to detect these regressions. To achieve this, we manually inject performance regressions into the programs, execute and collect new trace data, and evaluate whether we can leverage the performance model to detect the injected performance regressions. Our approach for detecting performance regressions is illustrated in Figure 3.

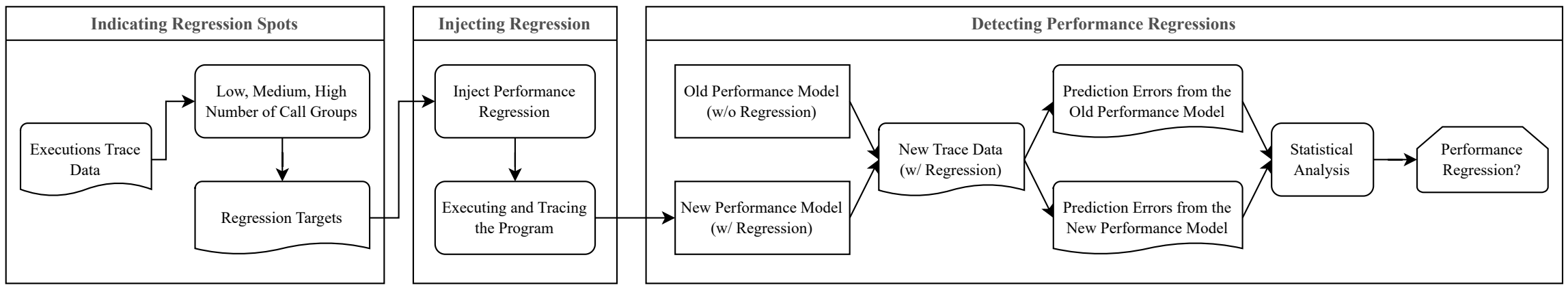

Fig. 3. The general overview of detecting performance regressions with the optimized performance model.

#### 3.4.1 *Choosing Performance Regression Injection Spots.*

To inject performance regressions into the programs, we face the challenge of choosing suitable target functions. Although regressions can occur in any function, injecting regressions into all functions and evaluating our regression detection method is impractical and infeasible. To address this, we developed an approach for selecting representative code locations for regression injection. First, we cluster all functions—both performance-sensitive and performance-insensitive—based on their average call frequency (calculated from the analysis dataset) into three groups: low, medium, and high call frequencies.

The clustering process begins by removing outliers from the distribution of functions' average call frequency using the z-score metric. We then determine the first quartile (Q1) and the third quartile (Q3) of the distribution. Functions with an average call frequency below Q1 are classified as "low-call," those between Q1 and Q3 as "medium-call," and those above Q3 as "high-call." From each cluster, we randomly select a number of functions (i.e., five) to form the set of target functions. As a result, we create three groups of functions—five from each cluster, totaling 15 functions—marked as targets for performance regression injection.

It is important to note that we inject only one regression at a time, affecting a single function in each trace data collection process. This allows us to accurately assess the effectiveness of the optimized performance model in detecting regressions when only one function is impacted.

*3.4.2* ***Injecting Performance Regressions and Collecting New Data.*** In this study, as the performance model predicts the program's total execution time, we focus on injecting a code block that introduces a time delay within the function's execution, resulting in an increase in the function's self-execution time and a decrease in CPU utilization due to the absence of any meaningful processing. After conducting investigations into the previously used performance regression types [4, 43, 51], we identified this regression type as the most suitable for our study. In other words, injecting additional types of performance regressions (e.g., involving more calculations or I/O operations) would not yield significantly more valuable insights in this context. After each regression injection, we rebuild the programs and execute them with new previously unseen inputs (the same number of executions we use to build the optimized performance models). The collected trace data from the new versions of the programs will be used to evaluate the optimized performance models' efficiency in detecting performance regressions.

*3.4.3* ***Detecting Performance Regressions.*** To detect performance regressions between two versions of a program (with and without the regression), we follow a methodology similar to that used in previous research on performance models [43]. We compare the predictions generated by performance models for both versions, allowing us to identify potential regressions.

Building performance models for each program version. For both versions of the program, we build separate performance models using optimized trace data. The optimization criterion (e.g., *Entropy*, *CoV*, *Performance Correlations*) is chosen based on a comparison of model accuracy and tracing overhead for each criterion. This ensures that the most effective criterion is used for constructing the performance models for both versions.

Comparing performance models to detect regressions. After building the models for the old (without regression) and new (with regression) program versions, we compare their predictions on the same new dataset. Our hypothesis is that performance regressions can be detected by examining differences between the two models' predictions. Specifically, we expect that any significant variation in model predictions between versions signals a potential performance regression. Following previous research [43], we employ the *Mann–Whitney U Test*[46] and *Cliff's Delta Effect Size*[18] to evaluate the differences in predictions. The *Mann–Whitney U Test* assesses whether the differences in predictions between the two models are statistically significant, while *Cliff's Delta* quantifies the magnitude of these differences. We choose these non-parametric tests because they do not require a normal data distribution. If the statistical difference ($p\text{-}value<0.05$) and effect size (e.g., at least a small effect) meet our predefined thresholds, we can conclude that the new version exhibits a performance regression or improvement.

Baseline: Direct comparison of performance before and after changes. To validate the effectiveness of our performance models in identifying regressions, we establish a baseline approach where we directly compare the system's performance before and after the changes (i.e., regression injections). In this baseline, we apply the *Mann–Whitney U Test* and *Cliff's Delta Effect Size* directly to the execution times from both versions. This is a common method for detecting performance regressions [36, 43]. Finally, we compare the results of our performance model approach against those obtained through this baseline, assessing the relative effectiveness of both methods.

### 3.5 Application Scenarios

Our approach is centered on performance modeling and regression detection within continuous integration and production environments. The process may be done in an automated pipeline, which involves the following steps:

(1) *Identifying performance-sensitive functions:* In continuous integration environments, full tracing (with higher overhead) is conducted less frequently—typically once per release. This tracing identifies performance-sensitive functions using our dynamic approaches. Alternatively, the static method can be used to identify these functions without introducing any additional overhead, bypassing the need for full tracing.
(2) *Applying optimized tracing:* Once performance-sensitive functions are identified, either through dynamic or static methods, optimized tracing (with lower overhead) is applied more frequently, usually for each commit. This step is crucial for building performance models and detecting regressions.
(3) *Monitoring in production:* Due to its minimal overhead, the optimized tracing can also be deployed in production environments to detect regressions under real-world conditions.

By following this approach, we ensure effective performance modeling and regression detection with minimal impact on system resources.

## 4 EVALUATIONS

### 4.1 Experiment Setup

The programs under study were compiled and executed on a machine equipped with an Intel Core i7-11700K processor (3.60 GHz), 16GB of RAM, and a 1TB SSD NVMe drive. For tracing, we used uftrace version 0.14[2], and the trace data was stored in a local database for later analysis.

**Trace data collection for identifying performance-sensitive functions.** Each program was executed 2,500 times with varying inputs (see Section 3.4.3) to capture a wide range of call paths. In practical applications, this data collection process can be time-consuming. However, the identification of performance-sensitive functions (based on a threshold, see Section 3.2.3) is independent of other functions. Therefore, this step can be performed incrementally. For example, when new functions are added to the codebase, only the new functions need to be tested to collect their trace data without requiring a full re-run of all tests. Also, as discussed in the *Application Scenario* (see Section 3.4.3), this trace collection process may be performed at the release level, which typically happens infrequently.

**Trace data for performance modeling and regression detection.** To efficiently construct performance models for each new program version, we use a smaller, statistically representative sample of the trace data collected during the identification of performance-sensitive functions. We select this sample with a 95% confidence level and a 5% margin of error, resulting in trace data from 333 execution inputs. This ensures that the sample adequately represents the dataset without requiring the full 2,500 inputs.

To prevent potential overlap with the trace data used in previous performance models (e.g., from different versions of the program), we prepare new, unique sets of execution inputs for each model. This helps reduce the risk of identical performance behaviors, ensuring that the performance models can effectively differentiate between versions.

To evaluate the accuracy of the criteria used to identify performance-sensitive functions (see Section 3.2.2), we collect optimized tracing datasets for each criterion and for both the original program version and the version with performance regressions.

[2] https://github.com/namhyung/uftrace/releases/v0.14

Table 2. Number of traced functions in programs for each tracing criterion

| Program | Full | Dynamic | | | | | | | | | Static |
|---|---|---|---|---|---|---|---|---|---|---|---|
| | | Entropy | | CoV | | Entropy & CoV | | Perf. Corr. | Feat. Sig. | All | SPS |
| | | w/o CR | w/ CR | w/o CR | w/ CR | w/o CR | w/ CR | | | | |
| SU2 | 745 | 42 (5.6%) | 4 (0.5%) | 13 (1.7%) | **2 (0.3%)** | 48 (6.4%) | 6 (0.8%) | 5 (0.7%) | 9 (1.2%) | 12 (1.6%) | 0 (0.0%) |
| 638.imagick | 418 | 46 (11.0%) | 7 (1.7%) | 27 (6.5%) | 7 (1.7%) | 61 (14.6%) | 14 (3.4%) | 12 (2.9%) | **4 (1.0%)** | 16 (3.8%) | 24 (5.7%) |
| 657.xz | 114 | 25 (21.9%) | **2 (1.8%)** | 9 (7.9%) | **2 (1.8%)** | 29 (25.4%) | 4 (3.5%) | **2 (1.8%)** | 0 (0.0%) | 4 (3.5%) | 26 (22.8%) |
| 631.deepsjeng | 98 | 15 (15.3%) | **2 (2.0%)** | 14 (14.3%) | **2 (2.0%)** | 29 (29.6%) | 4 (4.1%) | 9 (9.2%) | 13 (13.3%) | 16 (16.3%) | 34 (34.7%) |
| freqmine | 34 | 7 (20.6%) | 3 (8.8%) | 10 (29.4%) | **2 (5.9%)** | 16 (47.1%) | 5 (14.7%) | 5 (14.7%) | 13 (38.2%) | 14 (41.2%) | 13 (38.2%) |

CR = Correlation Removal (removes highly correlated functions). Full = tracing all executed functions without optimizations.
SPS = StaPerfSens. Bold values indicate the smallest percentage for each program.

### 4.2 RQ1: How does tracing only performance-sensitive functions impact tracing overhead?

Tracing can introduce substantial performance overhead [6, 30, 42, 59], limiting its use in performance-sensitive environments such as production systems with heavy user interaction. In this research question, we investigate how focusing on tracing only performance-sensitive functions can reduce this overhead.

To address RQ1, we prepared the programs with their respective inputs and conducted tracing using two approaches: full tracing and optimized tracing based on different criteria for identifying performance-sensitive functions (see Section 3.2.2). Table 2 provides details on the number of functions traced for each program, while Table 3 summarizes the resulting overhead in terms of execution time and storage usage.

**Only a small subset of functions in each program are performance-sensitive.** As discussed earlier, not all functions contribute equally to overall system performance [6, 33, 59]. Using performance-insensitive function pruning (see Section 3.2.2), we found that only a small portion of functions significantly impact performance. These are the performance-sensitive functions we target for tracing when building performance models. Table 2 shows that only a fraction of functions need to be traced, depending on the criterion. For instance, using the *Entropy-based* pruning criterion, only 5.64% to 21.93% of functions require tracing. Similarly, the *CoV-based* criterion traces between 1.74% and 29.41% of functions. The union of these two sets expands the traced functions to 6.44% to 47.06%. Notably, correlation removal further reduces these numbers, leading to tracing as little as 0.54% to 8.82% of functions.

**Tracing only performance-sensitive functions significantly reduces overhead.** Table 3 demonstrates the reduction in execution time and storage usage achieved by pruning functions compared to full tracing. Full tracing incurs substantial execution time overhead, ranging from 58.04% to 2243.88%, and consumes between 4.26MB and 3.92GB of storage. By contrast, *Entropy-based* pruning reduces execution time overhead to between 31.34% and 2098.07%, and storage usage to 3.42MB to 1.67GB. Even greater reductions are achieved with *CoV-based* pruning, which brings execution time overhead down to between 4.59% and 177.78%, and storage usage to between 2.60MB and 515.92MB. Additionally, other criteria, such as *Performance Correlations* and *Feature Significance*, further reduce overhead. For example, execution time overheads with *Correlation-based* pruning range from 11.47% to 39.66%, and storage usage is reduced to between 3.13MB and 159.76MB.

**No single criterion fits all programs; different criteria produce different sets of performance-sensitive functions.** Although *Entropy* and *CoV* both measure function stability, their results often differ. As explained in Section 3.2.2, *Entropy* captures randomness in function execution times, while *CoV* measures variability in those times. This difference means the two criteria identify distinct sets of performance-sensitive functions. For example, Table 2

Table 3. The average tracing overhead in each program, including execution time and storage usage

| Program | Full | Dynamic | | | | | | | | | Static |
|---|---|---|---|---|---|---|---|---|---|---|---|
| | | Entropy | | CoV | | Entropy & CoV | | Perf. Corr. | Feat. Sig. | All | SPS |
| | | w/o CR | w/ CR | w/o CR | w/ CR | w/o CR | w/ CR | | | | |
| **Average Execution Time Overhead (Compared to Vanilla Execution in Percentage)** | | | | | | | | | | | |
| SU2 | 77.1% | 34.7% | 12.5% | 4.6% | **2.2%** | 35.4% | 14.3% | 12.5% | 4.0% | 12.3% | 0.0% |
| 638.imagick | 168.4% | 100.4% | 25.0% | 28.3% | **14.3%** | 103.1% | 32.0% | 28.3% | 23.1% | 28.8% | 23.4% |
| 657.xz | 2243.9% | 2098.1% | 31.0% | 177.8% | 29.3% | 2222.2% | 45.2% | 28.4% | 0.0% | 43.2% | **19.5%** |
| 631.deepsjeng | 471.9% | 31.3% | **0.4%** | 57.0% | 5.0% | 85.9% | 7.3% | 11.5% | 110.8% | 115.0% | 57.9% |
| freqmine | 58.0% | 44.2% | 47.8% | 55.2% | 48.4% | 50.9% | 52.2% | **39.7%** | 48.3% | 50.0% | 57.2% |
| **Average Storage Usage Overhead** | | | | | | | | | | | |
| SU2 | 748.6M | 487.0M | 143.5M | 23.3M | **7.4M** | 501.0M | 147.3M | 140.9M | 21.6M | 158.8M | 0.0B |
| 638.imagick | 899.4M | 832.4M | 137.7M | 144.1M | **30.2M** | 876.5M | 166.7M | 159.8M | 114.9M | 166.7M | 42.3M |
| 657.xz | 1.79G | 1.67G | 14.2M | 140.2M | 13.3M | 1.79G | 26.4M | 14.6M | 0.0B | 26.4M | **7.8M** |
| 631.deepsjeng | 3.92G | 285.4M | **1.0M** | 515.9M | 34.6M | 800.3M | 34.7M | 81.8M | 976.8M | 977.0M | 560.2M |
| freqmine | 4.3M | 3.4M | **1.8M** | 2.6M | 1.8M | 4.2M | 2.1M | 3.1M | 3.7M | 4.0M | 2.4M |

CR = Correlation Removal (removes highly correlated functions). SPS = StaPerfSens. Bold values indicate the smallest overhead for each program. In the storage overhead comparison section, M is for megabytes, and G is for gigabytes.
StaPerfSens for SU2 and Feature Significance for 657.xz did not provide candidate performance-sensitive functions, hence their overheads are zero.

shows that *Entropy-based* pruning traces 5.64% to 21.93% of functions, while *CoV-based* pruning traces 1.74% to 29.41%. However, the overlap between these sets is minimal. For the programs studied, the number of common functions traced by both criteria is small (e.g., 6, 12, 5, 0, and 1 for *SU2*, *638.imagick*, *631.deepsjeng*, *657.xz*, and *freqmine*, respectively).

**Removing correlations between functions enhances overhead reduction.** Initially, *Entropy-based* and *CoV-based* pruning identified performance-sensitive functions based on individual function characteristics. However, functions often exhibit interdependencies, leading to redundancy in the data. To address this, we applied the *Spearman* correlation method [56] to identify highly correlated functions ($\rho \geq 0.7$) and merged them using hierarchical dendrograms. Removing highly correlated functions further reduced the number of functions requiring tracing. With *Entropy-based* pruning, only 0.54% to 8.82% of functions needed to be traced after correlation removal, and *CoV-based* pruning required tracing only 0.27% to 5.88%. The union of the two sets resulted in 0.81% to 14.71% of functions being traced. This refinement led to significantly lower execution time overhead, ranging from 0.45% to 47.76%, and storage usage reduced to between 1.02MB and 166.58MB.

**RQ1 Conclusion**

Both dynamic and static approaches for pruning performance-insensitive functions can lead to a substantial reduction in tracing overhead. However, there are no universally best-performing pruning criteria for tracing overhead reduction across different programs. Also, considering the inter-correlations between the performance of the functions can further help reduce the tracing overhead.

Table 4. The accuracies of the best optimized performance model of each program compared to full tracing performance models

| Program | Best Approach | | Optimized Performance Model | | | Full Tracing Performance Model | | |
|---|---|---|---|---|---|---|---|---|
| | **Pruning Criterion** | **Performance Model** | **MAE** | **RMSE** | **R2** | **MAE** | **RMSE** | **R2** |
| SU2 | Entropy and CoV (w/o CR) | CatBoostRegressor | 0.145 | 0.213 | 0.998 | 0.146 | 0.231 | 0.998 |
| 638.imagick | Feature Significance | CatBoostRegressor | 0.039 | 0.065 | 0.997 | 0.065 | 0.108 | 0.996 |
| 657.xz | StaPerfSens | BayesianRidge | **0.002** | **0.003** | **0.999** | 0.571 | 1.240 | 0.950 |
| 631.deepsjeng | StaPerfSens | BayesianRidge | **0.031** | **0.053** | **0.999** | 2.186 | 3.579 | 0.960 |
| freqmine | Performance Correlations | AdaBoostRegressor | 0.004 | 0.005 | **0.931** | 0.004 | 0.006 | 0.916 |

### 4.3 *RQ2: How well can we build performance models with only performance-sensitive functions?*

Accurate performance models are crucial for various tasks, such as system comprehension [19, 34, 63], capacity planning [10, 37], and performance regression detection [43, 44, 51]. In this study, we aim to develop performance models that rely solely on tracing information from performance-sensitive functions (i.e., with minimal overhead). These models are designed to be applicable in resource-constrained or performance-sensitive environments, such as production systems. Using the methods described in Section 3.2, we constructed optimized performance models based on execution frequencies of performance-sensitive functions (see Section 3.3) for each pruning criterion. The accuracy of these models was then compared to a baseline model built with full tracing data.

**Although no single pruning method consistently leads to the best performance model, performance-sensitive functions can accurately explain program performance variations.** Table 4 presents the highest accuracies achieved for each program and pruning approach (e.g., using the *Feature Significance* set for *638.imagick*). The optimized performance models, constructed from performance-sensitive functions, showed impressive accuracy with $R^2$ values ranging from 0.931 to 0.999, sometimes surpassing models built from full tracing data. While no single pruning approach consistently yielded the best results, *StaPerfSens* proved to be particularly effective, enabling optimized models to achieve the highest $R^2$ scores. Additionally, the *CatBoostRegressor* [49] and *BayesianRidge* models emerged as the top performers across various programs. Considering all evaluation metrics (MAE, RMSE, and $R^2$), the *StaPerfSens* criterion outperformed others in building highly accurate performance models for *657.xz* and *631.deepsjeng*. In contrast, for *SU2*, the *Entropy and CoV (without correlation removal)* criterion produced the best results. For *638.imagick* and *freqmine*, the *Feature Significance* and *Performance Correlations* criteria led to better model accuracy than other pruning methods.

**The number of traced functions does not always correlate with performance model accuracy.** As mentioned in previous sections, each pruning criterion generates a different set of performance-sensitive functions, leading to variations in model accuracy. The results show that the top-performing optimized models frequently outperformed those built from full tracing data, suggesting that model accuracy is not necessarily tied to the number of traced functions. This discrepancy can be explained by the high correlation among many functions in terms of their performance behavior. For example, in the case of *SU2*, the *Entropy and CoV (without correlation removal)* set traces 48 functions, while the full tracing set includes 745 functions. Despite containing far fewer functions, the optimized model built from the reduced set (using *CatBoostRegressor*) demonstrated slightly better accuracy than the full tracing model.

**Choosing the right model type is critical to effectively leverage performance-sensitive functions.** In our experiments, we used *PyCaret* to train, tune, and test various performance models. The most effective model type varied across programs, with *CatBoostRegressor* and *BayesianRidge* often providing the best results. For instance, *BayesianRidge* performed best for *657.xz* and *631.deepsjeng*, while *AdaBoostRegressor* was the optimal choice for *freqmine*.

**RQ2 Conclusion**

Optimized performance models built with performance-sensitive functions are capable of accurately capturing the performance of the programs, even outperforming the *full tracing* performance models in some cases. However, it is crucial to select the optimal pruning criterion and performance model for the subject programs.

### 4.4 RQ3: How effective are the optimized performance models in detecting performance regressions?

In this research question, we explore whether performance models built with optimized tracing data can effectively detect performance regressions. To ensure a representative analysis, we identified three clusters of functions based on their call frequency: low, medium, and high. A delay (e.g., 5 µs) was injected into these functions one at a time, and the programs with the regression were executed with new inputs. Having 15 new versions for each program, fresh performance trace data was collected for the performance-sensitive functions in each version. To determine the most suitable pruning criterion (e.g., *Entropy* or *CoV*) for building performance models of the new versions (i.e., with regression), we considered three factors: (1) the criterion that produced the highest model accuracy (i.e., highest $R^2$ score); (2) the one that reduced tracing overhead the most (i.e., minimized execution time); and (3) the one that achieved the best balance between accuracy improvement (compared to full tracing) and overhead reduction. Table 5 presents the performance model that delivered the best results for each program. Detailed information from this table is provided in our replication package (see 3.4.3).

**Optimized performance models can distinguish between program versions with and without regressions, outperforming the classical approach of directly comparing performance.** Table 5 shows the results of comparing performance models for regression detection. Differences between models built from two versions indicate performance variations (i.e., regressions). A version without regression serves as the baseline, and the goal is to detect regressions without false positives. For instance, in *631.deepsjeng*, when no regression is present, the p-value of the model built using *Entropy and CoV (without correlation removal)* exceeds 0.05, with negligible effect size, indicating no performance regression. However, for the 15 versions with injected regressions, p-values consistently fall below 0.05, with medium to large effect sizes observed, signaling performance regressions. By setting a threshold for the p-value and effect size, we can accurately detect regressions. The effectiveness of optimized performance models, compared to direct execution time comparison, is evident in Table 5. For example, in *SU2*, although versions without regression showed insignificant p-values and negligible effect sizes, the baseline approach (i.e., comparing execution times) failed to detect regressions in the modified versions. In contrast, the optimized models identified regressions across all programs, while direct comparison failed for programs like *638.imagick*, *631.deepsjeng*, and *freqmine*. This shows the importance of performance models, especially when workloads vary over time, as simple execution time comparisons often fail in such cases.

**A performance model's accuracy metrics (e.g., $R^2$) are closely tied to its effectiveness in detecting regressions.** When evaluating the effectiveness of performance models in distinguishing between versions with and without regressions, metrics such as MAE, RMSE, and $R^2$ play a significant role. A model with higher accuracy is generally more effective in detecting performance regressions. As discussed in the results of RQ2, choosing the optimal combination of pruning criteria (e.g., *Entropy* or *CoV*) and performance model types (e.g., *LinearRegression* or *AdaBoostRegressor*) is critical, as their accuracy varies across scenarios. In this research question, we observe that if an optimized model performed better in RQ2, it also showed superior regression detection capabilities. The full results of the performance regression comparisons are available in our replication package (see 3.4.3).

Table 5. The effectiveness of the optimized performance models in detecting performance regressions compared to the direct comparison of execution times

**SU2**

| Approach | | Without Regression | | With Regression | | | |
|---|---|---|---|---|---|---|---|
| | | **P-value** | **Effect Size** | **P>0.05 or ES=N** | **ES=S** | **ES=M** | **ES=L** |
| Entropy and CoV (w/o CR) | Performance Model | **0.600** | **L. [0.579]** | **0/15** | **0/15** | **4/15** | **11/15** |
| | Direct Comparison | 0.946 | N. [-0.107] | 14/15 | 1/15 | 0/15 | 0/15 |

**638.imagick_s**

| Approach | | Without Regression | | With Regression | | | |
|---|---|---|---|---|---|---|---|
| | | **P-value** | **Effect Size** | **P>0.05 or ES=N** | **ES=S** | **ES=M** | **ES=L** |
| Feature Significance | Performance Model | **0.077** | **S. [0.177]** | **0/15** | **4/15** | **6/15** | **5/15** |
| | Direct Comparison | «0.01 | S. [-0.164] | 7/15 | 3/15 | 2/15 | 4/15 |

**631.deepsjeng_s**

| Approach | | Without Regression | | With Regression | | | |
|---|---|---|---|---|---|---|---|
| | | **P-value** | **Effect Size** | **P>0.05 or ES=N** | **ES=S** | **ES=M** | **ES=L** |
| Entropy (w/ CR) | Performance Model | **0.845** | **N. [-0.020]** | **0/15** | **0/15** | **2/15** | **13/15** |
| | Direct Comparison | «0.01 | M. [0.406] | 6/15 | 4/15 | 4/15 | 1/15 |

**657.xz_s**

| Approach | | Without Regression | | With Regression | | | |
|---|---|---|---|---|---|---|---|
| | | **P-value** | **Effect Size** | **P>0.05 or ES=N** | **ES=S** | **ES=M** | **ES=L** |
| StaPerfSens | Performance Model | **0.325** | **N. [0.099]** | **0/15** | **1/15** | **0/15** | **14/15** |
| | Direct Comparison | 0.167 | N. [0.062] | 6/15 | 1/15 | 1/15 | 7/15 |

**freqmine**

| Approach | | Without Regression | | With Regression | | | |
|---|---|---|---|---|---|---|---|
| | | **P-value** | **Effect Size** | **P>0.05 or ES=N** | **ES=S** | **ES=M** | **ES=L** |
| Performance Correlations | Performance Model | **0.073** | **S. [0.180]** | **1/14** | **0/14** | **0/14** | **13/14** |
| | Direct Comparison | «0.01 | S. [-0.177] | 0/15 | 1/15 | 3/15 | 10/15 |

*Note:* ES = Effect Size (calculated using Cliff's Effect Size). N = Negligible, S = Small, M = Medium, L = Large. Results shown are for the most accurate performance models. Complete results for all models are available in the replication package.

**RQ3 Conclusion**

Our optimized performance models have shown the ability to distinguish between program versions with and without performance regressions, surpassing the baseline of purely comparing execution times without using performance models. This underscores the importance of employing performance models in scenarios where workloads evolve over time, enabling the differentiation between versions affected by regressions and those that are not. Also, the most accurate performance models demonstrate the highest capability in detecting performance regressions.

# 5 THREATS TO VALIDITY

## 5.1 External Validity.

This study examined five well-developed programs renowned for CPU-intensive tasks and extensively used in prior analyses. While these programs are predominantly written in C and C++, it's essential to highlight the broader applicability of our approach beyond these languages. The principles of information gain and function fluctuations are universal across programming languages, extending our method's potential to Java or Python programs. We employed *uftrace* for tracing, specifically designed for C/C++/Rust program execution tracing. While alternative instrumentation tools like *LTTng* exist, it's notable that their impact on system overhead may differ. However, the reduction in overhead remains consistent across all tools, given the identical number of functions removed from instrumentation.

### 5.2 Internal Validity

The programs were executed individually on a single workstation without parallelism to prevent resource bottlenecks. While external interruptions were minimized during executions, internal system interruptions might have influenced function behavior fluctuations. Outlier trace data was removed during feature engineering, but some inaccurate trace information may still be present. For example, in *freqmine*, despite fewer functions traced in *Entropy with correlation removal* compared to *Entropy without correlation removal*, its tracing overhead (i.e., execution time) was higher. This discrepancy could be attributed to unforeseen system interruptions during program execution.

### 5.3 Construct Validity

We identified performance-sensitive functions by analyzing the trace data collected from program executions using diverse inputs. As it is impractical to cover every possible input scenario for a specific program, the inputs used in the experiments aimed to range from low to high workloads. Yet, certain functions might exhibit uncovered behavior with unseen inputs, potentially influencing the identification of performance-sensitive functions. More diverse inputs (e.g., through fuzzing) could be considered in future work to identify sensitive functions.

In this study, we used *Shannon's Entropy* and *Coefficient of Variation* as the variance-based analysis, and the *Performance Correlations* and *Feature Significance* method as the model-impact analysis to determine the performance-sensitive and performance-insensitive functions. However, since not all of the program's trace data would align perfectly with a single common performance model (e.g., *RandomForestRegressor* or *CatBoostRegressor*), we chose not to explore additional metrics. Nevertheless, in the pursuit of deploying an ideal optimized performance model, it is worthwhile to explore various metrics for the trace data to identify the most appropriate set of performance-sensitive functions.

## 6 CONCLUSIONS

We propose an automated approach that leverages optimized tracing to construct performance models and detect performance regressions. We find that only a small portion (about 5% on average) of the functions of the studied programs are performance-sensitive. Such performance-sensitive functions can be automatically identified by different criteria (e.g., based on the fluctuation of the performance of the functions or their contribution to the performance models). By tracing only these performance-sensitive functions, we can build accurate and lightweight performance models at a minimum tracing overhead (just about 80% reduction on average compared to baseline models). Our experiment results also demonstrate that such lightweight performance models can be leveraged to detect performance regressions. With the low overhead and the automated process, our approach can be used to automatically identify optimal tracing code spots (i.e., performance-sensitive functions), build performance models, and detect performance regressions in a production environment where the overhead is sensitive. Our work also sheds light on future work that leverages optimized tracing and lightweight performance models in other performance engineering tasks.

## 7 DATA AVAILABILITY

We share a replication package to encourage future work to replicate or build on our work.[3].

[3] https://github.com/mooselab/replication-package-tracing-optimization-performance-modeling